\documentclass[twocolumn,english,aps,prx,floatfix,amssymb,superscriptaddress,longbibliography]{revtex4}
\usepackage[latin9]{inputenc}
\usepackage{verbatim}
\usepackage{float}
\usepackage{amsmath}
\usepackage{amssymb}
\usepackage{mathtools}
\usepackage{graphicx}
\usepackage{color}
\usepackage{xcolor}
\usepackage{tikz}
\usepackage[normalem]{ulem}
\newcommand{\ket}[1]{\ensuremath{\left| #1 \right>}}

\usepackage{bbold}

\usepackage[bookmarks=false,linkcolor=blue,urlcolor=blue,colorlinks,citecolor=blue]{hyperref}
\makeatletter

\newcommand{\be}{\begin{equation}}
\newcommand{\ee}{\end{equation}}
\newcommand{\bea}{\begin{eqnarray}}
\newcommand{\eea}{\end{eqnarray}}

\begin{document}
\title{Realizing a Symmetry Protected Topological Phase in a Superconducting Circuit}

\author{Parameshwar R. Pasnoori}
\affiliation{Department of Physics, University of Maryland, College Park, MD 20742, United
States of America}
\affiliation{Laboratory for Physical Sciences, 8050 Greenmead Dr, College Park, MD 20740,
United States of America}
\author{Patrick Azaria}
\affiliation{Laboratoire de Physique Th\'eorique de la Mati\`ere Condens\'ee, Sorbonne Universit\'e and CNRS, 4 Place Jussieu, 75252 Paris, France}
\author{Ari Mizel}
\affiliation{Laboratory for Physical Sciences, 8050 Greenmead Dr, College Park, MD 20740,
United States of America}

\begin{abstract}
We propose a superconducting quantum circuit whose low-energy degrees of freedom are described by the sine-Gordon (SG) quantum field theory. For suitably chosen parameters,
 the circuit hosts a symmetry protected topological (SPT) phase  protected by a discrete $\mathbb{Z}_2$ symmetry. 
 The ground state of the system is twofold degenerate and exhibits local spontaneous symmetry breaking of the  $\mathbb{Z}_2$ symmetry close to the edges of the circuit, leading to spontaneous localized edge  supercurrents. The ground states host Majorana zero modes (MZM) at the edges of the circuit. On top of each of the two ground states, the system exhibits localized bound states at both edges, which are topologically protected against small disorder in the bulk.  The spectrum of these boundary excitations should be observable in a circuit-QED experiment with feasible parameter choices.
\end{abstract}
\maketitle

Symmetry protected topological phases (SPTs) exhibit non-local order parameters, transcending Landau's symmetry-breaking characterization of phase transitions.  Moreover, SPT phases possess zero-energy edge modes which enjoy robustness against local disorder \cite{NayakMZM,Alicea_2012}. As a result of these intriguing features, SPT phases have commanded substantial scientific attention \cite{AKLT,HALDANE1983,Keselman,Boulat,chen2017flux,Diehl2015,Keselman2018,Starykh2000,Ruhman2012,Zoller2013,Starykh2007,nijs,pollman,Wen}.  One well-known example of an SPT arises in certain one-dimensional superconductors \cite{Alicea_2012}.   In these systems, superconductivity is induced by proximity effects, and they exhibit Majorana zero modes (MZM) \cite{Kitaev_2001,Kanemaj}.
Recently, it was shown that SPTs can also occur in one-dimensional quantum wires where superconductivity is induced by intrinsic attractive interactions among electrons \cite{Keselman,PAA1,PAA2,Parmeshthesis}.  Several other proposals to realize MZM have been suggested \cite{Yazdani2021,yuvalmaj,Saumaj,dasmaj}, and  observations have been reported \cite{mannamzm,Tanakamzm,fanmzm,Agheemaj,gerboldmaj}, but realizing and manipulating them remains extremely challenging.  In this article, we propose a superconducting quantum
circuit that realizes an SPT and its associated MZM, leveraging a technology over which refined experimental control
has already been achieved \cite{cqed1,circuitqed,transmon,PhysRevB.77.180502,clarke,ManucharyanFluxonium,RevModPhys.93.025005}. 

\begin{figure}
\includegraphics[width=1\columnwidth]{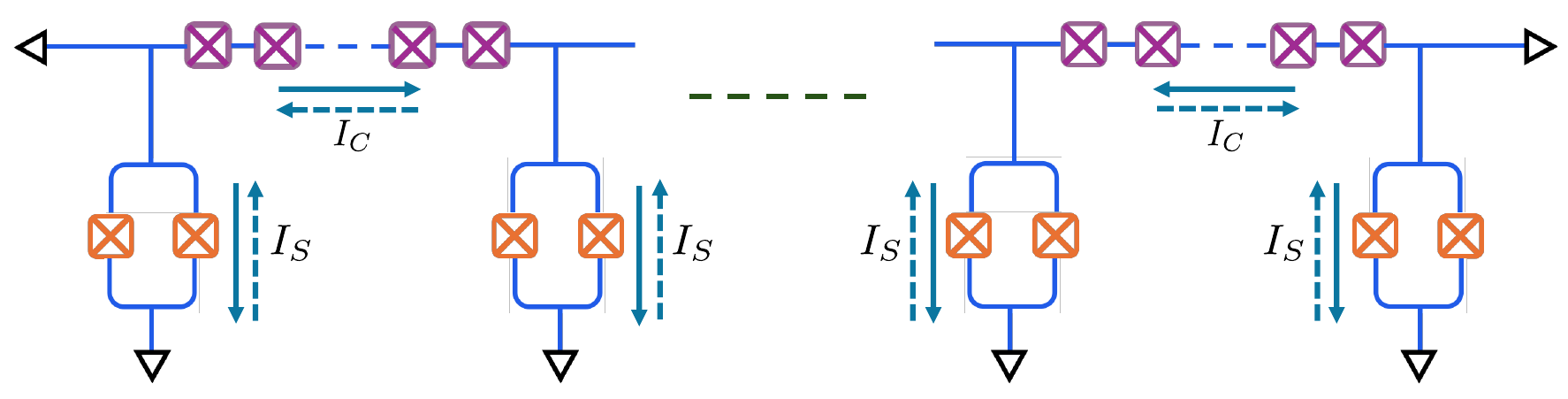}
\caption{Proposed circuit that exhibits the SPT phase of the SG model. The circuit consists of $M+1$ SQUIDs (orange) connected to each other through couplers. Each coupler contains $N$ Josephson junctions (purple) in series. The bosonic fields of the SG model correspond to the phase drops across the SQUIDs. When the two ends of the circuit are grounded as shown, and one superconducting flux quantum is threaded through each SQUID, the circuit exhibits two degenerate ground states. In both ground states, supercurrents flow across the couplers ($I_C$) and the SQUIDs ($I_S$). The direction of the current flow in one ground state is indicated by solid arrows. In the other ground state it is equal and opposite and is indicated by dashed arrows.}   
\label{circuittop}
\end{figure}


Our circuit's low-energy degrees of freedom are described by the sine-Gordon quantum field theory.
Using Bethe ansatz 
\cite{PRD}, we show that this theory exhibits an SPT phase with a twofold degenerate ground state hosting two MZM.  The SPT phase is protected by a discrete $\mathbb{Z}_2$ symmetry that is spontaneously broken locally at the edges, giving rise to spontaneous supercurrents. 
On top of the ground states, the system exhibits a rich spectrum of excitations localized at the boundary which are topologically protected against small local disorder.  These excitations provide a spectroscopic signature of the SPT phase accessible to circuit-QED measurements.

\paragraph{Quantum circuit.}
We consider the Josephson-junction array 
\cite{LarkinSC,kondocircuit,RoySG,solitoncircuit}
depicted in Fig. \ref{circuittop}.  The $M+1$ vertical links of the circuit are formed by SQUIDs.  The effective capacitance of each SQUID is $c_b$ and one superconducting flux quantum, $h/2e$, is threaded through each SQUID such  that the effective junction energy $E_J^b$ is {\it negative}. We denote the phase drop across each SQUID by $\Phi_k$, where $k=0,\dots,M$.  
To couple $\Phi_k$ to its neighbor $\Phi_{k+1}$, the circuit includes a horizontal array of $N$ Josephson junctions, each with Josephson energy $E_J^a$ and capacitance $c_a$.  
The phase drop across the $i^{th}$ junction of the $l^{th}$ coupler is denoted by $\Theta_{N(l-1)+i}$, where $l = 1,\dots, M$ and $i= 1,\dots, N$.   
The Lagrangian of the circuit can be written as 
\cite{yurke,yurke:87,PhysRevA.43.6414,1997devoret}
\begin{align}\nonumber
\mathcal{L}=&\sum_{k=0}^{M} \frac{c_b}{2}\dot{\Phi}_k^2+ \sum_{l=1}^{M}\sum_{i=1}^{N}\frac{c_a}{2}\dot{\Theta}_{N(l-1)+i}^2\\ -
&\sum_{k=0}^{M}E_{J}^{b}(1-\cos\Phi_k)-\sum_{l=1}^{M}\sum_{i=1}^{N}E_{J}^a(1-\cos\Theta_{N(l-1)+i})\label{bareCLag}
\end{align} 
in units in which $\hbar/2e = 1$.
The first two terms correspond to the capacitance energies and the last two terms correspond to the Josephson energies of the circuit elements.
We ground the two ends of the circuit such that the phase drops across the SQUIDs at the two boundaries are zero: 
\be
\Phi(0)=\Phi(L)=0. \label{BC}
\ee

In order to obtain the  field theory that describes the low-energy physics associated with our circuit, we first assume that the couplers act as inductors in which the non-linearity of the cosine potential of the junctions corresponding to the couplers can be neglected \cite{ManucharyanFluxonium}.  Then, we integrate out the phase drops $\Theta_{N(l-1)+i}$ across the junctions in the couplers to obtain an effective Lagrangian governing just the $\Phi_k$. 
This is physically reasonable provided that excitations of the $\Theta_{N(l-1)+i}$ degrees of freedom are high in energy, which holds in the parameter regime 
\be\; |E_J^b| \ll E_J^a, \;\;  E_C^a \ll E_J^a, \; \frac{|E_J^b|}{ E_J^a}\ll \frac{N}{N-1} \frac{E_C^a}{ E_C^b},\label{derivativelimit}\ee 
where $E_C^{a,b}=e^2/(2c_{a,b})$.
As explained in detail in the Supplementary Material (SM), integrating out the $\Theta_{N(l-1)+i}$ fields inevitably  produces non-local interactions among the $\Phi_k$ arising from inverting the capacitance matrix.
This can be avoided  by working in the large $N$ limit with
\be E_C^a \gg \frac{E_C^b}{N-1}.\label{spacelimit}\ee
Finally, we take the continuum limit with $\Phi(x=k a_0)\equiv \Phi_k$ and $\sum_k = a_0^{-1} \int_0^{L} dx$, where $L=Ma_0$ and
 $a_0$ is the distance between the SQUIDs.  Eq. (\ref{bareCLag}) becomes the Lagrangian of the sine-Gordon field theory $L_{SG}=\int_0^L  dx \: \mathcal{L}_{SG}$, where
 
\begin{align}
 {\cal L}_{SG}= \frac{1}{8\pi K}\left( \frac{1}{u} (\partial_{t} \Phi)^2  -u (\partial_{x} \Phi)^2\right)+ \lambda\: (\cos\Phi-1).
\label{SGLagrangian}
\end{align}
Here, the dimensionless velocity scale $u$, the phase stiffness $K$ and the parameter $\lambda$ are given by
\be
u=a_0\sqrt{\frac{2 E_J^a E_C^b}{N}},\; K=\frac{1}{4 \pi}\sqrt{ \frac{2N E_C^b}{E_J^a}},\; \lambda=E_J^b{a_0}^{-1}\label{relCSG}.
\ee

The Lagrangian  (\ref{SGLagrangian}) is invariant under the $\mathbb{Z}_2$ symmetry \footnote{This is termed charge conjugation symmetry in standard treatments of the sine-Gordon field theory, but we avoid mentioning charge here to prevent confusion with electrical charge.}
\be
{\cal C}: \Phi(x) \rightarrow - \Phi(x)
\label{z2}
\ee
and conserves the total phase per unit flux quantum 
\be
\varphi= \int_0^L dx\; \varphi (x),\text{ where } \varphi (x) = \frac{1}{2\pi} \partial_x \Phi.
\label{phasedensity}
\ee
Notice that our normalization is fixed such that the presence of a soliton or an anti-soliton changes $\varphi$ by $\pm 1$.  In our circuit,  due to the boundary conditions (\ref{BC}), $\varphi$ is zero. When periodic boundary conditions are considered, a gap opens in the spectrum 
 when $0\le K \le 2$,
and the ground state is non-degenerate for all values of $\lambda$.
For the Dirichlet boundary conditions (\ref{BC}) the situation changes, and the physics crucially depends on the sign of $\lambda$. When $\lambda > 0$ the ground state is  non-degenerate and  describes a  ``trivial" phase. In sharp contrast, when $\lambda < 0$, the Lagrangian (\ref{SGLagrangian}) describes an SPT phase as we shall see.  For the boundary conditions (\ref{BC}), this SPT phase exhibits  a twofold degenerate ground state hosting MZM localized at the two edges of the system.  Hence, threading  one superconducting flux quantum  through each  SQUID in our circuit (\ref{circuittop}), which  makes $E_J^b <0$ and $\lambda <0$, allows one to  realize  an  SPT phase.

\begin{center}
\begin{figure}[!h]
\includegraphics[width=1\columnwidth]{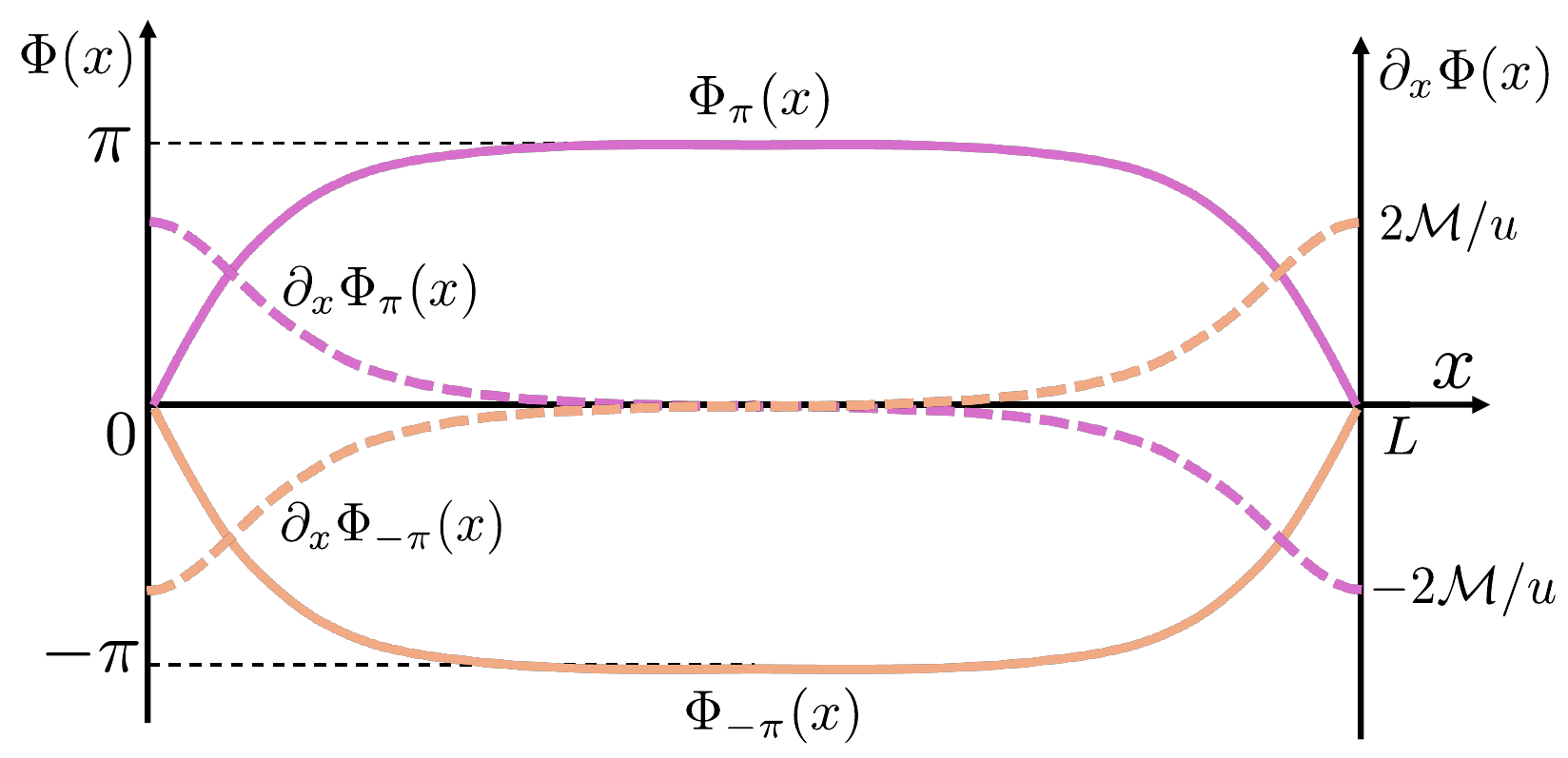}
\caption{Phases $\Phi_{\pi}(x)$ (solid purple), $\Phi_{-\pi}(x)$ (solid orange) and phase densities $\varphi_{L}(x)=\partial_x\Phi_{\pi}(x)$ (dashed purple), $\partial_x\Phi_{-\pi}(x)$ (dashed orange) corresponding to the two degenerate ground states in the semi-classical limit.}
\label{phases}
\end{figure}
\end{center}

\paragraph{Ground states and MZM.}
We find that in the thermodynamic limit $L\rightarrow \infty$, the ground state is twofold degenerate (to exponential accuracy in the system size).
The ground-state subspace is spanned by two $\varphi=0$ states 
\begin{equation}
   |0_{L}\rangle \text{ and } |0_{R}\rangle ,
 \label{GS}
\end{equation}
which transform into each other under   $\mathcal{C}$ (\ref{z2}), i.e: $\mathcal{C} |0_L\rangle =  |0_R\rangle$. Hence the symmetry (\ref{z2})
is spontaneously broken in the topological phase ($\lambda <0)$  in contrast with the trivial phase ($\lambda >0$) where the ground state is unique.
To build intuition about the nature of the ground states, consider the semi-classical limit obtained when $K\rightarrow 0$. Then the potential energy term in (\ref{SGLagrangian}), $-\lambda \cos \Phi$,  dominates in the Hamiltonian and the phase is fixed in the bulk at one of its minima $\Phi=(2n+1)\pi$, $n\in \mathbb{Z}$. With the boundary conditions (\ref{BC}), the phase field configurations  of minimal energy   are static solutions,  $\Phi_{\pm \pi}(x)$,  of the  equation of motion associated with (\ref{SGLagrangian}) that 
 interpolate between $\Phi=0$ at the edges $x=(0,L)$ and  $\Phi=\pi$ or $\Phi=-\pi$ in the bulk. As seen in Fig. \ref{phases},
 the two ground states display, in the semi-classical limit,
 non-zero phase density   profiles $\partial_x \Phi_{\pm \pi}$ close to the boundaries. Due to the gap in the bulk, the non-zero profiles are exponentially localized at the left and right edges. The two ground states $|0_{L(R)}\rangle$ defined in (\ref{GS})  correspond to the solutions $\Phi_{\pm \pi}(x)$; the subscript notation $L(R)$ was chosen for reasons that will become clear (see Eq. (\ref{fractional}) below).   Away from the semi-classical limit, the above considerations carry over to the average phase densities in the two ground-states
\begin{equation}
\varphi_{\alpha}(x)= \;\langle 0_{\alpha}|\varphi(x)|0_{\alpha}\rangle,\;  \alpha=L,R.
 \label{GSdensities}
 \end{equation}
An involved Bethe ansatz analysis \cite{PRD} reveals that $\varphi_{L(R)}(x)$  takes the  form
\begin{eqnarray}
\varphi_{L(R)}(x)=& \pm& \frac{1}{2} ( \Delta \varphi_{L}(x) - \Delta \varphi_{R}(x)), 
 \label{GSphaseaccu}
 \end{eqnarray}
where $\Delta \varphi_{L(R)}(x)$ are functions that are exponentially localized near the left and right edges and are  normalized  as 
\begin{eqnarray}
\int_{0}^{L} dx \; \Delta \varphi_{L(R)}(x) =1.
\label{norm}
 \end{eqnarray}
Obtaining the exact analytical form of these functions for general values of $K$ is a formidable task.  However, analytical expressions can be obtained in the semi-classical limit $K \rightarrow 0$ and at the free fermion point $K=1$. We find that 
\begin{align}
K\rightarrow 0: \hspace{0.1in} & \Delta \varphi_{L}(x) =\frac{2 \mathcal{M}}{\pi u \cosh( \mathcal{M} x/u)  }, \nonumber \\
K=1: \hspace{0.1in} & \Delta \varphi_{L}(x) =\frac{2 m}{u} e^{-2 m x/u },
 \label{GSdensities}
 \end{align}
 with  $\Delta \varphi_{R}(x) = \Delta \varphi_{L}(L-x) $.
 In the above equations, $\mathcal{M}=\sqrt{2 E_C^b |E_J^b|}$ is the fundamental breather mass in the semi-classical limit and $m=\pi |E_J^b|$ is  the soliton mass at $K=1$ (see SM). Expressions (\ref{GSphaseaccu}), (\ref{GSdensities})  show that in each of the ground states (\ref{GS}), non-zero phases accumulate at the boundaries despite the fact that the total phase $\varphi$ in (\ref{phasedensity}) is zero. Indeed, following Jackiw et al. \cite{Jackiw}, we consider the   phase accumulation  operators at each edge  $\sigma_{L,R}$ 
  \begin{eqnarray}
\sigma_{L}&=& \lim_{\eta \rightarrow 0}  \lim_{L \rightarrow \infty} \int_0^L dx\; e^{-\eta x} \varphi(x), \nonumber \\
\sigma_{R}&=& \lim_{\eta \rightarrow 0}  \lim_{L \rightarrow \infty}  \int_0^L dx\;  e^{-\eta(L- x)} \varphi(x),
 \label{sigma}
 \end{eqnarray}
where  the exponential $\eta$-filters are present to pick out the contributions of  $\Delta \varphi_{L(R)} (x)$ at the two edges \footnote{Notice the importance of order of limits in (\ref{sigma}). If the $\eta \rightarrow 0$ limit is taken first the operators  $\sigma_{L(R)}$ would identify with the total phase $\varphi$ which is zero}. 
Using (\ref{norm}) we find, for all values of $\lambda$ and $0<K<2$,
\begin{align}
\begin{array}{ll}
\langle 0_{L}|\sigma_{L}|0_{L}\rangle =  +\frac{1}{2}, &\hspace{0.1in} \langle 0_{L}|\sigma_{R}|0_{L}\rangle = - \frac{1}{2}, \\
& \\
\langle 0_{R}|\sigma_{R}|0_{R}\rangle = +\frac{1}{2},  & \hspace{0.1in}\langle 0_{R}|\sigma_{L}|0_{R}\rangle = -\frac{1}{2}, 
\end{array}
\label{fractional}
\end{align}
and  verify that    $\varphi= \langle 0_{\alpha}|\sigma_{L}|0_{\alpha}\rangle+\langle 0_{\alpha}|\sigma_{R}|0_{\alpha}\rangle= 0$. Thus, in the two ground-states (\ref{GS}), $\pm$ half a flux quantum  is trapped at the edges corresponding to  a trapped half soliton or anti-soliton. Accordingly
we may describe the ground state manifold \footnote{ It should be noted that there  exist consistently two other  states with total phase $\varphi=\pm 1$ obtained by acting with $\tau^x_{{L}}$
 and $\tau^x_{{R}}$ on $\ket{0}_{L}$ and $\ket{0}_{R}$ respectively. The resulting states
 \bea |\pm 1\rangle =|\pm \frac{1}{2}\rangle_{{L}} \otimes |\pm \frac{1}{2}\rangle_{{R}},  \label{tensorpm1}
\eea
are degenerate with  the  two ground states (\ref{tensor}) which  have total phase
 $\varphi=0$.  In terms of the circuit the states (\ref{tensorpm1}) can  be obtained by fixing the  boson fields at the boundaries as $\Phi(0)=0$ and $\Phi(L)=\pm 2 \pi$ $(\hspace{-0.1in}\mod 2\pi)$. In the Bethe ansatz  solution the operators $\tau^x_{{L(R)}}$ are zero energy modes related to the existence of boundary string solutions of the Bethe equations.} as
 \bea \ket{0_L} &\rightarrow& \ket{+\frac{1}{2}}_{{L}} \otimes \ket{-\frac{1}{2}}_{{R}},
 \nonumber \\
\ket{0_R} &\rightarrow& \ket{-\frac{1}{2}}_{{L}} \otimes\ket{+\frac{1}{2}}_{{R}}. \label{tensor}
\eea
Here the states $|\pm \frac{1}{2}\rangle_{L(R)}$  are eigenstates of the fractional  phase  operators (\ref{sigma})
\be
{\sigma}_{L(R)} |\pm \frac{1}{2}\rangle_{L(R)} =\pm \frac{1}{2} |\pm \frac{1}{2}\rangle_{L(R)}.
\label{edgestates}
\ee
The above description assumes that the fractional phase accumulations (\ref{edgestates}) are genuine quantum observables or equivalently that  the variances
$\delta \sigma^2_{L(R)}= \langle \sigma^2_{
L(R)} \rangle -  \langle \sigma^{}_{L(R)} \rangle^2 $
vanish in both ground states $\ket{0_{L(R)}}$. This is the case at the free fermion point $K=1$ \cite{Jackiw} and also in the semi-classical limit $K\rightarrow 0$ \cite{TakacsSC, MussardoSC}; we shall assume that this remains true for all $K$. As we shall now see, this implies the existence of 
MZM acting in the ground state manifold. Indeed, let us
define three Pauli matrices at each edge,  
$\tau^{\mu=x,y,z}_{L(R)}$, acting on the states (\ref{edgestates}) such that  $\tau_{L(R)}^z \equiv 2 \sigma_{L(R)}$.  Then,
$\tau_{L(R)}^x$ acts as local phase conjugation operator that transforms a trapped half soliton at the left (right) edge to an half anti-soliton and vice-versa.
Hence, when acting on the ground state manifold,  the  global phase conjugation $\mathbb{Z}_2$ symmetry (\ref{z2}) fractionalizes between the two edges  according to ${\cal C}=\tau_{L}^x\tau_{R}^x$. In this framework,  the zero energy Majorana fermion operators (MZM) are given by \cite{Fendley}
\begin{equation}
\xi_L= \tau_L^z,\hspace{0.1in} \xi_R= \tau_L^y \tau_R^x 
\end{equation}
such that
$\{\xi_r, \xi_{r'}\}=2 \delta_{r,r'}$
where $r,r'= L,R$ and
${\cal C}=i \xi_L \xi_R$.
In the basis (\ref{tensor}) the 
MZM $\xi_R$ transfers, up to a phase, a soliton from one edge to the other.
 
\begin{center}
\begin{figure}[!h]
\includegraphics[width=1\columnwidth]{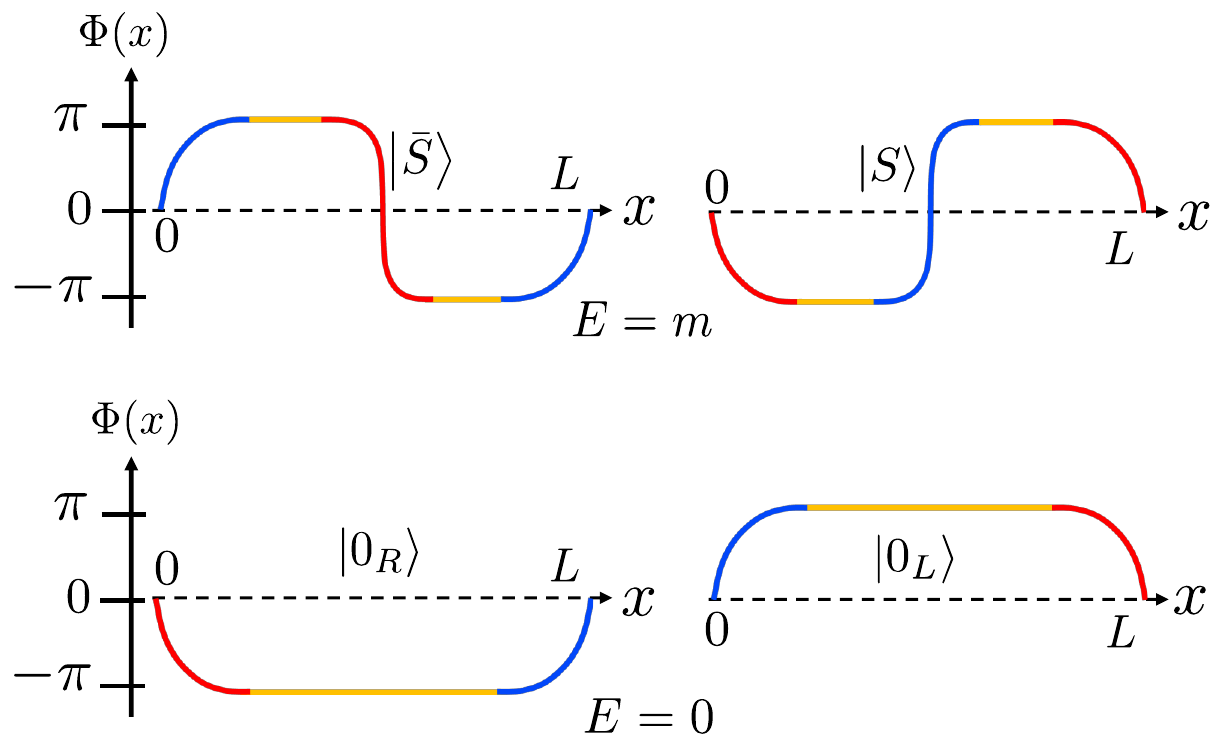}
\caption{ Pictorial representation of the phase field $\Phi(x)$ in the two ground-states $\ket{0_{L(R)}}$ and the two excited states $\ket{S}$ and  $\ket{\bar{S}}$. 
The state $\ket{S}$ contains a soliton (blue) and the state $\ket{\bar{S}}$  contains an anti-soliton (red) that propagate in the bulk and interpolate between the two bulk constant field configurations (orange). In all these states, a half soliton or half anti-soliton is trapped at each edge so that the total phase accumulation is $\varphi=0$ as required by the boundary conditions.}
\label{states}
\end{figure}
\end{center}

\paragraph{Excited states.}
In the repulsive regime, i.e. when $K\ge 1$, the first excited states are
 \begin{equation}
 |S\rangle \text{ and } |\bar S\rangle
 \label{Spm}
\end{equation}
with energy $E_{\theta}= m \cosh\theta$, where $m$ is the soliton mass and $\theta$ the rapidity.  These states are represented pictorially in Fig.\ref{states}.  They contain bulk solitons and anti-solitons, respectively,     
 which carry phase $\pm 1$.  This phase is compensated in $|S\rangle$ and $|\bar S\rangle$ by a localized half soliton/anti-soliton with phase accumulation $\mp 1/2$ trapped at each edge. 
In this way, the two states (\ref{Spm}) maintain total phase $\varphi = 0$ as required by the boundary conditions (\ref{BC}).  
 In the trivial phase, a single propagating soliton or an anti-soliton does not appear in the spectrum; instead the first excited state consists of a soliton$-$anti-soliton pair with gap $2m$. Hence the existence of the solitonic states (\ref{Spm}) is 
 one of the signatures of the topological phase. The whole spectrum is then obtained by adding to any of the four  states $|0_{L(R)}\rangle$, $|S\rangle$ and $|\bar S \rangle$, an arbitrary number of soliton$-$anti-soliton pairs.

\begin{center}
\begin{figure}[!h]
\includegraphics[width=0.8\columnwidth]{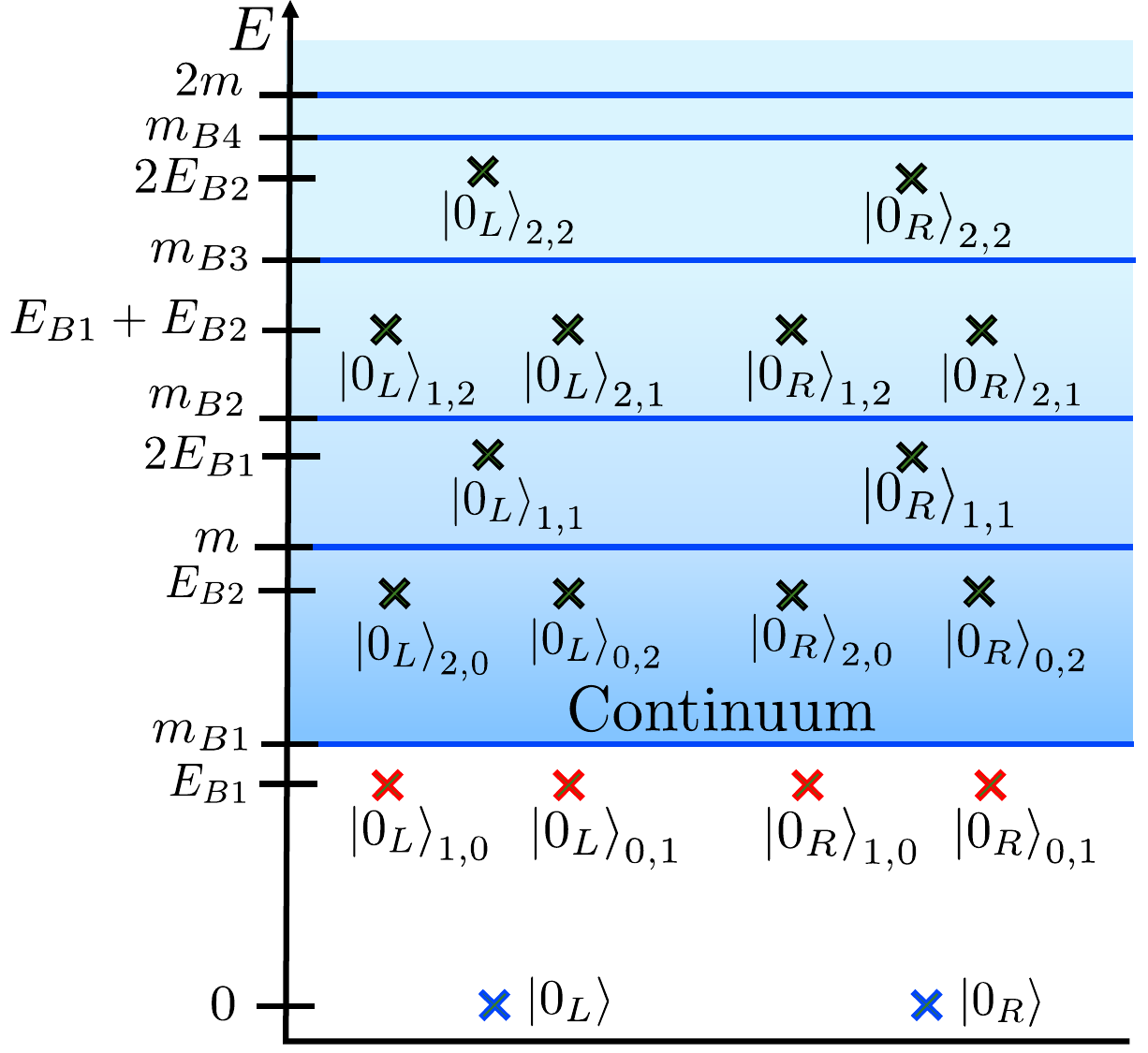}
\caption{Spectrum for $2/7<K\le2/5$ in the energy range $0\le E < 2m$.  There are two ground states $\ket{0_L}$ and $\ket{0_R}$ with energies set to zero (blue crosses). The blue region represents the continuum. Since $2/7<K\le2/5$, the continuum excitations include $4$ types of bulk breathers, with masses $m_{Bp}=2 m \sin \frac{p \pi \xi}{2}$,
$p=1,2,3,4$; the continuum starts at the mass of the lightest bulk breather $m_{B1}$.  Higher energy continuum excitations also include solitons and anti-solitons starting at mass $m$.   Sixteen boundary breather states are shown, four of which with energy $E_{B1}$ are ``mid-gap'' states below the continuum (red crosses).  The remaining twelve (black crosses) are above the continuum threshold with energies $E_{B2}$, $2E_{B1}$, $E_{B1}+E_{B2}$ and $2E_{B2}$.}

\label{towers}
\end{figure}
\end{center}

 In the attractive regime $K < 1$, both in the trivial and the topological phases, soliton and anti-soliton bind
 together to form bulk breathers  \cite{Thacker} with masses
\begin{equation}
m_{Bp}= 2 m \sin(\frac{p \pi}{2} \xi),\; p= 1,...,[\frac{1}{\xi}],\; \xi=\frac{K}{2-K},
\label{bulkbreathers}
\end{equation}
where $m$ is the soliton mass and  $[...]$ denotes the integer part.
In the topological phase in the region $\frac{2}{3} < K < 1$ the first excited states are $|S\rangle$,  $|\bar S\rangle$. The next excited states are obtained by adding the first bulk breather bound-state (i.e. $p=1$)  to either of the  $|0_{L(R)}\rangle$ ground states. For sufficiently large attractive interactions, i.e: when $K < \frac{2}{3}$,  in addition to the above excitations 
{\it boundary  breather} bound-states emerge. 
These states are localized at the left and/or the right boundaries and have energies
\begin{equation}
E_{Bp}=  m \sin(p \pi\xi),\; p= 1,...,[\frac{1}{2\xi}].
\label{boundarybreatherm}
\end{equation}
They correspond to a soliton (anti-soliton) bound to the fractional anti-soliton (soliton) that is trapped at an edge of the system. 
The net phase accumulation due to a  boundary breather (with respect to the ground states $\ket{0_{L(R)}}$) is zero. 
One can sort these eigenstates into degenerate  left and right towers (see Fig. \ref{towers})
\begin{eqnarray}
|0_L\rangle_{p,q}\; {\rm and}\; |0_R\rangle_{p,q},\hspace{0.2in} p,q=0,...,[\frac{1}{2\xi}],
\label{boundarybreatherstowers}
\end{eqnarray}
with the convention  $|0_{L(R)}\rangle_{0,0} \equiv |0_{L(R)}\rangle$.  The states $|0_{L(R)}\rangle_{p,q}$ are obtained after adding two boundary breathers on top of the ground state $|0_{L(R)}\rangle$: one   at the left boundary with energy $E_{Bp}$ and  one at the right boundary   with energy $E_{Bq}$. 
As a result, the energy of $|0_{L(R)}\rangle_{p,q}$ is  $E_{Bp}+ E_{Bq}$.  Notice that, with our notation, the states $|0_{L(R)}\rangle_{p,0}$ or $|0_{L(R)}\rangle_{0,p}$ contain only one boundary breather bound-state with energy $E_{Bp}$ at the left or the right edge respectively. These two towers (\ref{boundarybreatherstowers})  transform into each other under the $\mathbb{Z}_2$ symmetry, i.e: ${\cal C }|0_L\rangle_{p,q}= |0_R\rangle_{p,q}$.   The degeneracy in each tower is a consequence of space parity symmetry which exchanges $p$ and $q$. Fig.\ref{towers} depicts the spectrum 
in the topological phase for $\frac{2}{7} < K < \frac{2}{5}$.

Overall,  for $K>1/2$, the gap to the continuum is   the    soliton/anti-soliton mass $m$  and for $K< 1/2$ it is the mass of   the lightest bulk  breather $m_{B1}=2 m \sin(\frac{ \pi}{2} \xi)$. 
When  $K < \frac{2}{3} $ the first excited states are the four localized boundary breathers $|0_{L(R)}\rangle_{0,1}$ and $  |0_{L(R)}\rangle_{1,0}$ with energy $E_{B1}=m \sin( \pi\xi) < m_{B1}$ (\ref{boundarybreatherm}) which are thus mid-gap states. In this regime, the existence of these  mid-gap states is one of the  signatures of the topological phase.

\paragraph{Insulating phase instability.} 
It is well known that Josephson junction arrays are intrinsically unstable with respect to the formation of insulating phases \cite{LarkinSC}.  The instability, which results from commensurability effects with the underlying lattice, 
occurs at rational
mean Cooper-pair filling fractions $\rho_0 =p/n$ where $(p,n)$ are co-prime integers.  A  Mott phase is favored when $n=1$ and a Charge Density Wave (CDW) phase is favored when $n\ge 2$ \cite{Haldane_1981,BHphases, FisherMott, WhiteMott}. This is modeled by including the following perturbation in (\ref{SGLagrangian}) 
\be
{\cal L}_{SG} \rightarrow {\cal L}_{SG} - V  \cos{2n \Theta(x)}.
\label{HMott}
\ee
 Here, $\Theta$  is the dual field
of $\Phi$, i.e:
$[\Phi(x), \Theta(y)]=i \pi Y(x-y)$ with $Y(u)$ the Heaviside function.
Since the terms  $\cos \Phi(x)$ and $\cos 2n \Theta(x)$  are mutually non-local, they favor incompatible superconducting and  Mott/CDW orders. From the renormalization group point of view, the scaling dimensions of both cosine terms in (\ref{HMott}) are given by \cite{Giamarchi}
\bea
\Delta_{\rm Super}= K, \; \Delta_{\rm Mott/CDW}=\frac{n^2}{4 K},
\eea
respectively. We then expect superconducting order  when the Mott/CDW potential is the less relevant operator, i.e. when $\Delta_{\rm Super} < \Delta_{\rm Mott/CDW}$ or when $K < n/2$.  
Therefore, it is possible to avoid the insulating phases by working in the attractive regime with $K< 1/2$.

\paragraph{Edge supercurrent signature of the topological phase.}
As emphasized above, the ground states $\ket{0_{L(R)}}$ host a fractional soliton or anti-soliton localized at each edge of the system.  They give rise to supercurrents flowing at the edges of the circuit, an experimental signature of the SPT phase. Indeed, the currents through the SQUIDs and couplers are given by
\be
I_{S}(x)= E_J^b \sin\Phi(x),\; I_{C}(x) =\frac{u}{4 \pi K} \partial_x \Phi(x),
\label{squidscurrent}
\ee
since $u/( 4 \pi K a_0) = E_J^a/N$ is the inverse inductance of the couplers.
When averaged in the two ground states
$|0_{L(R)}\rangle$ the currents
$\langle I_S(x)\rangle_{L(R)}$ and 
$\langle I_C(x)\rangle_{L(R)}$
satisfy the equation of motion  of the sine-Gordon field theory
\be
a_0 \partial_x \langle I_{C}(x)\rangle_{L(R)} + \langle I_{S}(x)\rangle_{L(R)} =0,
\label{currcons}
\ee
which reflect the conservation of  total current at each node $x$ of the circuit.
Eventually the average currents through the SQUIDs and couplers can be expressed in terms of the phase densities (\ref{GSphaseaccu})
 \bea
\langle I_C(x)\rangle_{L(R)}&=& \pm \frac{u}{4K} \left( \Delta \varphi_{L} (x) - \Delta \varphi_{R} (x) \right),
\label{currents}
 \eea
whereas $\langle I_S(x)\rangle_{L(R)}$
is obtained from (\ref{currcons}). 
 In the $|0_{L}\rangle$ ground state, the current flows into the couplers
and exits to ground after flowing through the SQUIDs near the edges of the circuit. The direction of the current flow is opposite in the $|0_{R}\rangle$ ground state (see arrows in Fig. \ref{circuittop}).
Hence, both  phase conjugation and time reversal symetries are spontaneously broken in each of the 
$|0_{L(R)}\rangle$ ground states. 
It is  though important to note that these currents  are non-zero only close to the edges where $\varphi_{L(R)}(x) \neq 0$ (\ref{GSphaseaccu}).
This reflects the nature of the  topological phase where the $\mathbb{Z}_2$ symmetry (\ref{z2}) is  spontaneously broken only  ``locally" at the edges. The $\mathbb{Z}_2$ symmetry (\ref{z2}) remains  unbroken in the bulk where $\sin \Phi$ and  $\partial_x \Phi$ average to zero.

\paragraph{Spectroscopic signature of the topological phase.} By suitably coupling a resonator to the boundary circuit elements, one can employ circuit-QED methods to probe the spectrum associated with the boundary breather excitations. 
 In the regime $K<1/2$, which avoids the Mott/CDW phases as discussed above, the lowest energy excitations are four mid-gap boundary breathers with energy $E_{B1}$; the next excitations are bulk breathers with energy $m_{B1}$. 
We seek circuit parameters that allow the boundary breathers to be distinguished spectroscopically from the bulk breathers.
For numerical estimates of the energies (\ref{bulkbreathers}) and (\ref{boundarybreatherm}) of these excitations, we need
the soliton mass $m$. 
It can be expressed \cite{zamolo95} (see SM)  in terms of the circuit parameters as 

\begin{align}
\frac{m}{E_C^b}
= \left( \frac{|E_J^b|}{E_C^b}\right)^{\frac{1}{2-K}} \hspace{-0.2in}\frac{2 \Gamma(\frac{\xi}{2})}{(2 \pi K)^{\frac{1-K}{2-K}}\sqrt{\pi} \Gamma(\frac{1+\xi}{2})}\left( \frac{\pi \Gamma(1-\frac{K}{2})}{2\Gamma(\frac{K}{2})} \right)^{\frac{1}{2-K}}.
\label{Msolu}
\end{align}
For concreteness, we propose an experimental setup made of $L \sim 50000$ junctions consisting of $M+1=101$ SQUIDs connected with couplers made of $N=500$ Josephson junctions each.  
If we set $E_C^b=1$ GHz and $E_J^a= 100$ GHz, Eq. (\ref{relCSG}) gives $K \sim 0.25$. Taking $E_C^a$ roughly of order $1$ GHz, the constraints (\ref{derivativelimit}) are then satisfied provided $|E_J^b| \ll 100$ GHz.  Varying $|E_J^b|$ in the range $1-10$ GHz, we find that $E_{B1}= 1.49-5.58$ GHz,  $m_{B1}= 1.53-5.72$ GHz, and soliton mass $m=3.45-12.86$ GHz.
Thus, the mid-gap boundary breather states are separated from the bulk breather excitations by $\Delta =m_{B1}-E_{B1} \simeq 40-140$ MHz, which is sufficiently large to be resolved spectroscopically.
At this point it is worth noting the importance of the parameter $N$ in our circuit. Indeed, from Eq. (\ref{relCSG}) we see that lowering $N$ would decrease $K$ as small as $K\simeq 0.01$ for $N$ of order one and $E_J^a= 100$ GHz. In this case, one reaches the semiclassical limit in which both boundary and bulk breathers form a continuum. The gap $\Delta$ between the mid-gap bound-states and the first bulk breather becomes impractically small (of the order $\Delta \simeq 0.1$ MHz for $|E_J^b|=10$ GHz).

In this work, we proposed a superconducting quantum circuit whose low-energy degrees of freedom, the phase drops across the SQUIDs, are described by the sine-Gordon quantum field theory.
When one superconducting flux quantum is threaded through every SQUID, the system exhibits an SPT phase. In this phase, the system has a twofold degenerate ground state that hosts MZM localized at the boundaries. These MZM are associated with spontaneous supercurrents that flow across the circuit elements close to the boundaries. We showed that on top of the ground states, the system exhibits localized bound states at the boundaries which are topologically protected against small disorder and form a degenerate spectrum.
This boundary spectrum, which is one of the key signatures of the topological phase, can be probed using existing circuit-QED techniques.  
When the stiffness parameter $K$ is tuned such that it is not too close to zero,  the spectrum corresponding to the boundary breather excitations (\ref{boundarybreatherm}) is anharmonic.  This raises the interesting question as to whether our circuit 
could find use as a qubit; we hope to address this in the future.

\acknowledgments
We acknowledge fruitful discussions with K. Sardashti.  

\bibliography{circuit}

\appendix

\begin{widetext}

\section{Circuit Lagrangian} 
In this section we derive an effective Hamiltonian which describes the low energy degrees of freedom of the circuits shown in the figure (\ref{circuittop}). The circuit consists of $M+1$ SQUIDS attached to each other through couplers. The bottom ends of all the SQUIDS are connected to the ground.
 Each coupler contains $N$ Josephson junctions connected in series whose capacitances and Josephson junction energies are $c_a, E_J^a$ respectively. Flux threaded in the SQUIDs is chosen such that the effective junction energy is negative. The capacitances and Josephson junction energies of the SQUIDS are denoted by $c_b,E_J^b$ respectively. We ground the two ends of the circuit, which results in Dirichlet boundary conditions where the phase drop across the SQUIDS at the left and the right boundaries is

\be \Phi_0=0,\;\; \Phi_M=0. \label{dbclsup}\ee\vspace{5mm}

We will show in the following that in a certain regimes of the parameters associated with the circuit, the low energy degrees of freedom are the phase drops across the SQUIDS, and that the dynamics associated with these modes is governed by a Lagrangian which takes the form of the sine-Gordon model in the continuum limit.

\begin{center}
\begin{figure}[!h]
\includegraphics[width=0.8\columnwidth]{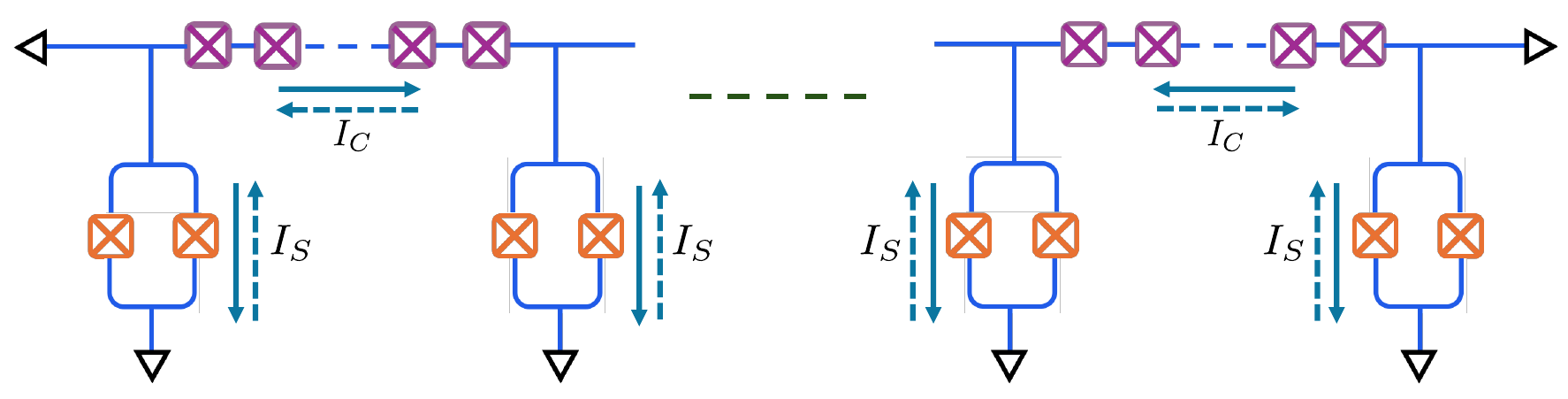}
\caption{Figure shows the circuit which corresponds to the sine-Gordon (SG) model. The circuit consists of $M+1$ SQUIDS attached to each other through couplers. Each coupler contains $N$ Josephson junctions connected in series whose capacitances and Josephson junction energies are $c_a, E_J^a$ respectively. The capacitances of the SQUIDS are denoted by $c_b$. The flux within each SQUID is tuned such that the effective Josephson energy $E^b_J$ is negative. The bosonic fields of the SG model correspond to the phase drops across the SQUIDS.}   
\label{circuittop}
\end{figure}
\end{center}

The phase drops across each SQUID is denoted by $\Phi_k$, where $k=0,1...M$. Let the phase drop across each junction in the $k^{th}$ coupler be denoted by $\Theta_{N(k-1)+i}$, where $i$ runs from $1$ to $N$, which is the number of junction in each coupler. The sum of the phase drops across all the junctions in any loop is zero. This gives rise to the relation

\bea \Theta_{Nk}=\Phi_k-\Phi_{k-1}-\sum_{i=1}^{N-1}\Theta_{N(k-1)+i}. \;\; (N>1)\label{looprule}
\eea

The Lagrangian of the circuit takes the form  \cite{1997devoret}

\begin{align} &\mathcal{L}=\nonumber\sum_{k=0}^{M} \frac{c_b}{2}\dot{\Phi}_k^2+ \sum_{l=1}^{M}\sum_{i=1}^{N-1}\frac{c_a}{2}\dot{\Theta}_{N(l-1)+i}^2+\sum_{l=1}^{M}\frac{c_a}{2}\left(\dot{\Phi}_{l}-\dot{\Phi}_{l-1}-\sum_{i=1}^{N-1}\dot{\Theta}_{N(l-1)+i}\right)^2\\
-\sum_{k=0}^{M}E_{J}^{b}(1-&\cos(\Phi_k))-\sum_{l=1}^{M}\sum_{i=1}^{N-1}E_{J}^a(1-\cos(\Theta_{N(l-1)+i}))-\sum_{l=1}^{M}E_{J}^a(1-\cos(\Phi_{l}-\Phi_{l-1}-\sum_{i=1}^{N-1}\Theta_{N(l-1)+i})).\end{align}

The first and the fourth terms are respectively the capacitance and Josephson energies associated with the SQUIDS. The second and the third terms are the capacitance energies associated with the couplers, whereas the fifth and sixth terms are the Josephson energies of the couplers. Note that we have used the relation (\ref{looprule}) to express the capacitance and Josephson energies of one of the junctions in each coupler in terms of the phase drops across the rest of the junctions and SQUIDS corresponding to the respective coupler. 

As we shall see, the phase drops across the SQUIDS correspond to the bosonic field in the sine-Gordon model. In order to obtain the kinetic term in the sine-Gordon model, it is necessary to take the limit $E_J^a \gg E_J^b, \; E_J^a\gg \frac{e^2}{2c_a}$ in the above equation. We obtain

\bea \mathcal{L}=\nonumber\sum_{k=0}^{M} \frac{c_b}{2}\dot{\Phi}_k^2+ \sum_{l=1}^{M}\sum_{i=1}^{N-1}\frac{c_a}{2}\dot{\Theta}_{N(l-1)+i}^2+\sum_{l=1}^{M}\frac{c_a}{2}\left(\dot{\Phi}_{l}-\dot{\Phi}_{l-1}-\sum_{i=1}^{N-1}\dot{\Theta}_{N(l-1)+i}\right)^2\\
-\sum_{k=0}^{M}E_{J}^{b}(1-\cos(\Phi_k))-\sum_{l=1}^{M}\sum_{i=1}^{N-1}\frac{E_{J}^a}{2}(\Theta_{N(l-1)+i})^2-\sum_{l=1}^{M}\frac{E_{J}^a}{2}(\Phi_{l}-\Phi_{l-1}-\sum_{i=1}^{N-1}\Theta_{N(l-1)+i})^2.\eea

The first line of the above equation which corresponds to the total capacitance energy of the circuit can be written as

\bea \frac{1}{2} (\begin{array}{cccccccc} \dot{\Phi}_{0}& \dot{\Phi}_{1}  & ....& \dot{\Phi}_{M}&\Theta_{1}&\Theta_{2}&...&\Theta_{NM-1} \end{array})\; \mathbb{C}\; \left(\begin{array}{c} \dot{\Phi}_{0}\\\dot{\Phi}_{1}\\\cdot\\\cdot\\\dot{\Phi}_{M}\\\Theta_1\\\Theta_2\\\cdot\\\cdot\\\Theta_{NM-1}\end{array}\right).\eea

For later convenience in notation, we express it in terms of the direct sum 

\bea \frac{1}{2}\bigoplus_{k=0}^{M-1} \left(\begin{array}{ccccccc}\dot{\Phi}_{k}&\dot{\Phi}_{k+1}&\dot{\Theta}_{Nk+1}&...&...&...&\dot{\Theta}_{N(k+1)-1}\end{array}\right)\mathbb{C}_k
\left(\begin{array}{c} \dot{\Phi}_{k}\\ \dot{\Phi}_{k+1}\\ \dot{\Theta}_{Nk+1}\\.\\.\\.\\\dot{\Theta}_{N(k+1)-1}\end{array}\right),\eea

where the capacitance matrix $\mathbb{C_k}$ is given by

\bea\nonumber \mathbb{C}_k= \left(\begin{array}{ccccccc}c_b&&&&&&\\&c_b&&&\\&&c_a&&&&\\ &&&c_a&&&\\ &&&&.&&\\ &&&&&.&\\&&&&&&c_a\end{array}\right)+
c_a\left(\begin{array}{c} \;\;\;1\\ -1\\ \;\;\;1\\.\\.\\.\\\;\;\;1 \end{array}\right)\left(\begin{array}{ccccccc} 1&-1&\;\;1&...&1&...&1\end{array}\right).\eea

We now apply a unitary transformation \bea U=\bigoplus_{k}^MU_k.\eea

 $U_k$ acts in the space spanned by $\Theta_{N(l-1+i)}$, $i=1,N-1$ for each $k \in (1,M)$, where 
\bea U_l=\left(\begin{array}{cccccc} \frac{1}{\sqrt{N-1}}&\frac{1}{\sqrt{N-1}}&......&\frac{1}{\sqrt{N-1}}&\frac{1}{\sqrt{N-1}}\\&&v_2&&\\&&v_3&&\\.\\.\\.\\&&v_{N-2}&&\end{array}\right),
\eea

where the row vectors $v_i, i\in (2,N-2)$ are orthogonal to each other and also orthogonal to the first row in $U_l$. Introducing new variables $\theta_{N(k-1)+i}=\sum_{j=1}^{N-1}U_k^{ij}\Theta_{N(k-1)+j}$, and noting that $\theta_{N(k-1)+1}=\frac{1}{\sqrt{N-1}}\sum_{i=1}^{N-1}\Theta_{N(K-1)+i}$, the Lagrangian takes the form 

\begin{align} \nonumber \mathcal{L}=\sum_{l=1}^M\sum_{i=2}^{N-1}\left(\frac{c_a}{2}\;\dot{\Theta}_{N(l-1)+i}^2-\frac{E_J^a}{2}\;\Theta_{N(l-1)+i}^2\right) &-\sum_{l=1}^M\frac{E_J^a}{2}\left(\Phi_{l}-\Phi_{l-1}-\sqrt{N-1}\theta_{N(l-1)+1}\right)^2\\\nonumber-\sum_{k=0}^{M-1} E_J^b\left(1-\cos(\Phi_k)\right)&-\sum_{l=1}^{M}\frac{E_J^a}{2}\theta_{N(l-1)+1}^2+L^{'},\end{align}

where \bea L^{'}=\bigoplus_{k=0}^{M-1}\left(\begin{array}{ccc}\dot{\Phi}_k&\dot{\Phi}_{k+1}&\dot{\theta}_{Nk+1}\end{array}\right) \mathbb{C}^{'}_k \left(\begin{array}{c} \dot{\Phi}_k\\\dot{\Phi}_{k+1}\\\dot{\theta}_{Nk+1}\end{array} \right), \;\; \mathbb{C}^{'}_k=\left(\begin{array}{ccc}c_b+c_a&-c_a&\sqrt{N-1}\;c_a\\-c_a&c_b+c_a&-\sqrt{N-1}\;c_a\\\sqrt{N-1}\;c_a&-\sqrt{N-1}\;c_a&c_a\end{array}\right). \eea

By making the transformation $\phi_{k+1}=\Phi_{k+1}-\Phi_{k}-\sqrt{N-1}\theta_{Nk+1}$, we have 

\bea\mathbb{C}^{'}_k=\left(\begin{array}{ccc}c_b+\frac{c_a}{N-1}&-\frac{c_a}{N-1}&-\frac{c_a}{N-1}\\-\frac{c_a}{N-1}&c_b+\frac{c_a}{N-1}&\frac{c_a}{N-1}\\-\frac{c_a}{N-1}&\frac{c_a}{N-1}&\frac{N c_a}{N-1}\end{array}\right). \eea

Non diagonal terms in the capacitance matrix $\mathbb{C}^{'}$ result in a Hamiltonian with interactions between spatially separated junctions. To avoid this, we need to work in the parameter regime where the capacitance matrix $\mathbb{C}^{'}$ takes a diagonal form. This corresponds to the limit

\bea c_b \gg \frac{c_a}{N-1}. \label{spatiallim}\eea

The Lagrangian now takes the form

\bea \nonumber &\mathcal{L}=\sum_{l=1}^M\sum_{i=2}^{N-1}\left(\frac{c_a}{2}\;\dot{\Theta}_{N(l-1)+i}^2-\frac{E_J^a}{2}\;\Theta_{N(l-1)+i}^2\right)\\\nonumber + \sum_{k=0}^{M}\frac{c_b}{2} \dot{\Phi}_k^2-&\sum_{k=0}^{M-1} E_J^b\left(1-\cos(\Phi_k)\right)+\sum_{k=0}^{M}\frac{c_a}{2} \dot{\phi}_{k+1}^2-\sum_{l=1}^{M}\frac{E_J^a}{2}\phi_{k+1}^2-\frac{E_J^a}{2(N-1)}\sum_{k=0}^{M-1}\left(\Phi_{k+1}-\Phi_{k}-\phi_{k+1}\right)^2.\eea

This can be written in a more compact way as

\bea \mathcal{L}= L_{\theta}+L_{\Phi}+L_{\phi},
\eea

where \bea L_{\theta}= \sum_{l=1}^M\sum_{i=2}^{N-1}\left(\frac{c_a}{2}\;\dot{\Theta}_{N(l-1)+i}^2-\frac{E_J^a}{2}\;\Theta_{N(l-1)+i}^2\right),\eea

\bea L_{\Phi}= \sum_{k=0}^{M}\frac{c_b}{2} \dot{\Phi}_k^2-\sum_{k=0}^{M-1} E_J^b\left(1-\cos(\Phi_k)\right)-\frac{E_J^a}{2(N-1)}\sum_{k=0}^{M-1}\Delta\Phi_{k+1}^2,\eea

and \bea L_{\phi}=\sum_{k=0}^{M-1} \frac{c_a}{2}\dot{\phi}^2_{k+1} -\frac{E_J^a}{2}\frac{N}{N-1}\phi_{k+1}^2 + \frac{E_J^a}{N-1}\Delta\Phi_{k+1}\phi_{k+1},
\eea

where $\Delta\Phi_{k+1}=\Phi_{k+1}-\Phi_k$. We see that the modes $\theta_{N(l-1)+i}$ which are described by the term $L_{\theta}$ are decoupled from the modes $\Phi_k$ and $\phi_{k+1}$ which are coupled to each other, and are described by the terms $L_{\Phi}+L_{\phi}$. To obtain an effective Lagrangian for the modes of interest $\Phi_k$, we need to integrate out the modes $\phi_{k+1}$ in $L_{\phi}$.  By following the usual method, we define functional integral for the modes $\phi_{k+1}$ 

\bea \int D\phi_{k+1}\;e^{i \int dt L_{\phi}}= \prod_{k=0}^{M-1}\int D\phi_{k+1}\; e^{\frac{i}{2}\int_0^T dt \;\phi_{k+1}(t)\left(-c_a\partial_t^2-E_J^{'}\right)\phi_{k+1}(t)}e^{i\int_0^T dt \;\alpha_{k+1}(t)\phi_{k+1}(t)},\eea

where \bea E_J^{'}= E_J^a \left(\frac{N}{N-1}\right), \; \alpha_{k+1}(t)=\left(\frac{E_J^a}{N-1}\right)\Delta\phi_{k+1}(t).\label{expalphaE}\eea 

Expanding the second exponential, we have  

\bea \nonumber \prod_{k=0}^{M-1}\int D\phi_{k+1}\; e^{\frac{i}{2}\int_0^T dt \;\phi_{k+1}(t)\left(-c_a\partial_t^2-E_J^{a'}\right)\phi_{k+1}(t)}\Huge(1+i\int_0^T dt \alpha_{k+1}(t)\phi_{k+1}(t)\\-\frac{1}{2}\int_0^T\int_0^T dt_1 dt_2 \alpha_{k+1}(t_1)\alpha_{k+1}(t_2)\phi_{k+1}(t_1)\phi_{k+1}(t_2)+...\huge).\label{pathint}\eea

Consider the first term

\bea Z=\nonumber \prod_{k=0}^{M-1}\int D\phi_{k+1}\; e^{\frac{i}{2}\int_0^T dt \;\phi_{k+1}(t)\left(-c_a\partial_t^2-E_J^{a'}\right)\phi_{k+1}(t)}.\eea

Taking the discrete Fourier transform \bea \phi_{k+1}(t) =\frac{1}{T} \sum_{n}e^{-i\omega_n t}\tilde{\phi}_{k+1}(\omega_n),\eea
 and working with the Euclidean coordinates $t\rightarrow-it^E$, $\omega_n\rightarrow i\omega^{E}_{n}$, we obtain

\bea  Z=\prod_{\omega^E_{n}}\left(\frac{-i\pi(L/c_a)}{(E_J^{'}/c_a)+(\omega^E_{n})^{2}}\right)^{M/2}.\label{Z}\eea

The second term in (\ref{pathint}) gives zero, whereas the third term gives

\bea \nonumber\prod_{k=0}^{M-1}-\frac{1}{2L^2}\int_0^T\int_0^T dt^E_{1}dt^E_{2} \alpha_{k+1}(t^E_{1})\alpha_{k+1}(t^E_{2})\sum_{\omega^E_{m}}e^{-i\omega^E_{m}(t^E_{1}-t^E_{2})} \prod_{\omega^E_n}\left(\frac{-i\pi(L/c_a)}{(E_J^{'}/c_a)+(\omega^E_{n})^{2}}\right)^{1/2}\left(\frac{-i\pi(L/c_a)}{(E_J^{'}/c_a)+(\omega^E_{m})^{2}}\right).\\\eea

Taking the continuum limit $\frac{1}{T}\sum_{\omega^E_m}=\int \frac{d\omega^E}{2\pi}$, we have

\bea \prod_{k=0}^{M-1}-\frac{1}{2c_a}\int_0^{\infty}\int_0^{\infty} dt^E_{1}dt^E_{2} \alpha_{k+1}(t^E_{1})\alpha_{k+1}(t^E_{2})\prod_{\omega^E_n}\left(\frac{-i\pi(L/c_a)}{(E_J^{'}/c_a)+(\omega^E_{n})^{2}}\right)^{1/2}\int_0^{\infty} \frac{d\omega^E}{2\pi} \left(\frac{e^{-i\omega^E(t^E_{1}-t^E_{2})}}{(E_J^{'}/c_a)+(\omega^E)^{2}}\right).\eea

Using the expression for $Z$ (\ref{Z}), and performing the integral over $\omega^E$ we obtain
\bea \prod_{k=0}^{M-1}-\frac{Z^{1/M}}{4\sqrt{E_J^{'a}c_a}}\int_0^{\infty}\int_0^{\infty} dt^E_1 dt^E_2 \alpha_{k+1}(t^E_1)\alpha_{k+1}(t^E_2)e^{-(E_J^{'}/c_a)^{1/2}|t^E_1-t^E_2|}.
\eea

In the limit
\bea E_J^b/c_b\ll E_J^{'}/c_a, \label{timelim}\eea

the fluctuations associated with the fields $\Phi_k$, and hence also that of $\alpha_k$, are negligible in the time scale $\Delta t \sim (E_J^{'}/c_a)^{-1/2}$ associated with the fluctuations corresponding to the fields $\phi_k$. In this limit, one can tailor expand $\alpha_{k+1}(t^E)$, which gives

\bea\prod_{k=0}^{M-1} -\frac{Z^{1/M}}{2\sqrt{E_J^{'a}c_a}}\int_0^{\infty}\int_0^{\infty} dt^E d(\Delta t) \alpha_{k+1}(t^E)\left(\alpha_{k+1}(t^E)+\Delta t \;\alpha_{k+1}^{'}(t^E)+...\right)e^{-(E_J^{'}/c_a)^{1/2}\Delta t}.
\eea

Keeping only the first order term, we have

\bea\prod_{k=0}^{M-1}-\frac{Z^{1/M}}{2\sqrt{E_J^{'a}c_a}}\int_0^{\infty}\int_0^{\infty} dt^E d(\Delta t) \alpha^2_{k+1}(t^E)e^{-(E_J^{'}/c_a)^{1/2}\Delta t}.
\eea

Hence we find that the limit (\ref{timelim}), the Lagrangian does not contain interactions between operators at unequal times. Evaluating the integral over $\Delta t$ and using the explicit expressions of $\alpha_{k+1}(t^E)$ and $E_J^{'}$ (\ref{expalphaE}), we obtain 
\bea\nonumber  \prod_{k=0}^{M-1}\int D\phi_{k+1}\; e^{\frac{i}{2}\int dt \;\phi_{k+1}(t)\left(-c_a\partial_t^2-E_J^{a'}\right)\phi_{k+1}(t)}\left(-\frac{1}{2}\right)\int dt_1 dt_2 \alpha_{k+1}(t_1)\alpha_{k+1}(t_2)\phi_{k+1}(t_1)\phi_{k+1}(t_2)\\=Z\prod_{k=0}^{M-1}-\frac{i E_J^{a}}{2N(N-1)}\int_0^{\infty} dt \Delta\Phi_{k+1}^2(t).\eea

Similarly, one can evaluate the higher order terms. We obtain

\bea  \prod_{k=0}^{M-1}\int D\phi_{k+1}\;e^{i \int dt L_{\phi}}= Z\prod_{k=0}^{M-1}e^{i\int dt L_{eff}}, \;\; L_{eff}=\frac{E_J^a}{2N(N-1)}\Delta\Phi_{k+1}^2(t).\eea

Combing $L_{eff}$ with $L_{\Phi}$, we obtain the effective Lagrangian governing the modes $\Phi_k$

\bea L_{SG}=\frac{c_b}{2}\sum_{k=0}^M \dot{\Phi}_k^2-\frac{E_J^a}{2N}\sum_{k=0}^{M-1}\Delta\Phi_{k+1}^2-E_J^b\sum_{k=0}^M(1-\cos\Phi_k).\eea


Taking the continuum limit 
\bea\Phi_k\equiv \Phi(ka_0)\;\;\; (x=ka_0),\\
\lim_{a_0\rightarrow 0}\sum_k a_0 \;\Phi(k a_0)=\int_0^L dx\; \Phi(x),
\eea

we obtain the following Lagrangian

\begin{align}
 L_{SG}= \int_0^L dx\left(  \frac{1}{8\pi K}\left( \frac{1}{u} (\partial_{t} \Phi)^2  -u (\partial_{x} \Phi)^2\right)+ \lambda(\cos\Phi-1) \right),
\label{SGLagrangian}
\end{align}
where, the dimensionless velocity scale $u$, the phase stiffness $K$ and the  parameter $\lambda$ are given by
\be
u=a_0\sqrt{\frac{2 E_J^a E_C^b}{N}},\; K=\frac{1}{4 \pi}\sqrt{ \frac{2N E_C^b}{E_J^a}},\; \lambda=E_J^b{a_0}^{-1}\label{relCSG}.
\ee

Here $E^{a,b}_{C}=1/2c_{a,b}$ is the capacitance energy. Let us recollect the limits (\ref{spatiallim}),(\ref{timelim}) in which this Larangian is valid:
 
 \be\; |E_J^b| \ll E_J^a, \;\;  E_C^a \ll E_J^a, \; \frac{|E_J^b|}{ E_J^a}\ll \frac{N}{N-1} \frac{E_C^a}{ E_C^b},\: E_C^a \gg \frac{E_C^b}{N-1}.\label{limitscollect}\ee

\section{Zamolochikov Mass Formula} 

Consider the sine-Gordon Lagrangian
\begin{equation}
L_{SG}=\int dx\; \left( \frac{1}{2}( \partial_{\mu} \phi)^2 + \lambda \:(\cos(\beta \phi)-1)\right),
\label{SSG}
\end{equation}
where the coupling $\lambda$ has the dimension of a [mass]$^2$, i.e: $\bar \lambda=\lambda/\Lambda^2$ is dimensionless, $\Lambda$ being an UV cutoff.
 Zamolochikov \cite{zamolo95} has obtained the exact expression for the soliton mass $m$  
\begin{equation}
m= \Lambda \frac{2 \Gamma(\frac{\xi}{2})}{\sqrt{\pi} \Gamma(\frac{1+\xi}{2})}\left( \frac{\pi \Gamma(1-\frac{K}{2})}{2\Gamma(\frac{K}{2})} \frac{|\lambda|}{\Lambda^2}\right)^{\frac{1}{2-K}},
\label{Msol}
\end{equation}
where
\begin{equation}
\xi=\frac{K}{2-K}.
\end{equation}
In our circuit we get, instead of (\ref{SSG}),   the Lagrangian
\begin{equation}
L_{SG}=\int dx\; \left( \frac{1}{2}( \frac{1}{u}(\partial_{t} \phi)^2 -  u (\partial_{x} \phi)^2 )+ \lambda \:(\cos(\beta \phi)-1)\right),
\label{CSG}
\end{equation}
where $\beta^2=4 \pi K$ and 
\begin{equation}
4\pi K= \sqrt{\frac{2E_C^b N}{E_J^a}}, \; u=\frac{E_C^b}{2 \pi K} a_0,\, \lambda = E_J^b a_0^{-1},
\label{couplings}
\end{equation}
  $a_0$ being the lattice spacing which plays the role of the UV cut-off. The circuit Lagrangian differs from (\ref{SSG}) by the dimensionless  velocity scale $u$. Using simple scaling arguments we find that
  \begin{equation}
m (\beta, \lambda,u)= u \; m (\beta, \frac{\lambda}{u},1),
\label{Msolu}
\end{equation}
where
\begin{equation}
\frac{\lambda}{u}= 2\pi K \frac{E_J^b}{E_C^b} a_0^{-2}.
\label{Msolu}
\end{equation}
We can now apply Zamolochikov formula (\ref{Msol}) by taking $\lambda/\Lambda^2=2\pi K \frac{E_J^b}{E_C^b} $ as the dimensionless coupling
at scale $\Lambda =a_0^{-1}$. The nice thing is that upon rescaling the soliton mass (\ref{Msol}) by $u$ the cut-off $a_0^{-1}$ dependence disappears in favor
of $E_C^b$. We eventually find
\begin{equation}
m(\beta, \lambda,u)= E_C^b \left( \frac{|E_J^b|}{E_C^b}\right)^{\frac{1}{2-K}}\; (\frac{1}{2 \pi K})^{\frac{1-K}{2-K} }\frac{2 \Gamma(\frac{\xi}{2})}{\sqrt{\pi} \Gamma(\frac{1+\xi}{2})}\left( \frac{\pi \Gamma(1-\frac{K}{2})}{2\Gamma(\frac{K}{2})} \right)^{\frac{1}{2-K}},
\label{Msolu}
\end{equation}
as quoted in the main text.
One may easily verify that at the free fermion point $K=1$ 
\begin{equation}
m=\pi |E_J^b|,
\end{equation}
while in the semi-classical limit $K\rightarrow 0$
\begin{equation}
m= \frac{\sqrt{8 E_C^b |E_J^b|}}{\pi K}.
\end{equation}
In the latter limit the spectral gap is given by the 
lightest bulk breather mass
which identifies with the energy of the first boundary breather $m_{B1}\simeq E_{B1}\equiv {\cal M}$ where
\begin{equation}
{\cal M}=\frac{\pi m K}{4}= \sqrt{2 E_C^b |E_J^b|}.
\end{equation}

\end{widetext}

\end{document}